\documentclass[acmsmall]{acmart}

\usepackage{xspace}
\usepackage{makecell}
\usepackage[breakable]{tcolorbox}

\AtBeginDocument{%
  }

\setcopyright{acmlicensed}
\copyrightyear{2018}
\acmYear{2018}
\acmDOI{XXXXXXX.XXXXXXX}

\acmConference[Conference acronym 'XX]{Make sure to enter the correct conference title from your rights confirmation emai}{June 03--05, 2018}{Woodstock, NY}
\acmISBN{978-1-4503-XXXX-X/18/06}

\begin{document}

\title{Python Symbolic Execution with LLM-powered Code Generation}

\author{Wenhan Wang}
\authornote{Both authors contributed equally to this research.}
\email{wenhan12@ualberta.ca}
\orcid{0000-0002-0585-2136}
\affiliation{%
  \institution{University of Alberta}
  \country{Canada}
}

\author{Kaibo Liu}
\authornotemark[1]
\affiliation{%
  \institution{Peking University}
  \country{China}
  }
\email{liukb@pku.edu.cn}

\author{An Ran Chen}
\affiliation{%
  \institution{University of Alberta}
  \country{Canada}
}
\email{anran6@ualberta.ca}

\author{Ge Li}
\affiliation{%
 \institution{Peking University}
 \country{China}
 }
 \email{lige@pku.edu.cn}

\author{Zhi Jin}
\affiliation{%
   \institution{Peking University}
 \country{China}
 }
 \email{zhijin@pku.edu.cn}

\author{Gang Huang}
\affiliation{%
 \institution{Peking University}
 \country{China}
 }
\email{hg@pku.edu.cn}

\author{Lei Ma}
\affiliation{%
  \institution{The University of Tokyo}
  \country{Japan}
}
\affiliation{%
  \institution{University of Alberta}
  \country{Canada}
}
\email{ma.lei@acm.org}

\newcommand{{\method}}[0]{LLM-Sym\xspace}

\begin{abstract}
Symbolic execution is a key technology in software testing, which generates test cases by collecting symbolic path constraints and then solving constraints with SMT solvers. Symbolic execution has been proven helpful in generating high-coverage test cases, but its limitations, e.g., the difficulties in solving path constraints, prevent it from broader usage in software testing. Moreover, symbolic execution has encountered many difficulties when applied to dynamically typed languages like Python, because it is extremely challenging to translate the flexible Python grammar into rigid solvers. 

To overcome the main challenges of applying symbolic execution in Python, we proposed an LLM-empowered agent, \method, that automatically calls an SMT solver, Z3, to solve execution path constraints. Based on an introductory-level symbolic execution engine, our LLM agent can extend it to supporting programs with complex data type `list'. The core contribution of \method is translating complex Python path constraints into Z3 code. To enable accurate path-to-Z3 translation, we design a multiple-step code generation pipeline including type inference, retrieval and self-refine. Our experiments demonstrate that \method is capable of solving path constraints on Leetcode problems with complicated control flows and list data structures, which is impossible for the backbone symbolic execution engine. Our approach paves the way for the combination of the generation ability of LLMs with the reasoning ability of symbolic solvers, and opens up new opportunities in LLM-augmented test case generation.
\end{abstract}



\maketitle

\section{Introduction}
Symbolic execution~\cite{king1976symbolic, baldoni2018survey} is a widely used program analysis technique which explores program execution paths with symbolic values. Together with constraint solving (with an SMT solver), symbolic execution can generate real variable values to satisfy a certain execution path. The key application of symbolic execution is software testing. Because symbolic execution can solve constraints for certain execution paths, we can use it to generate test cases that cover a specific execution path or expose hidden bugs.
Symbolic execution has been successfully applied to various program languages, including Java~\cite{galeotti2013improving, li2016symbolic}, LLVM IR~\cite{cadar2008klee} and binary code \cite{shoshitaishvili2016sok}.

However, symbolic execution faces significant difficulties when applied to dynamically typed languages such as Python \cite{bucur2014prototyping}. The major difficulty lies in mapping program execution paths to SMT solver constraints. First, without explicit type annotations, it can be difficult to define symbolic variables in the solver, where all symbolic variables must have their own data type. 
Second, existing symbolic execution engines often use rules to convert Python statements into SMT constraints \cite{zeller2019fuzzing, ball2015deconstructing}, but Python data structures are highly flexible. For example, a Python list can have an unfixed length, which is different from similar data structures in most other languages (e.g., array in C, Java). An obvious challenge this feature brings is that in Python programs, there are condition constraints on list lengths, and we may need to write extra SMT constraints to solve the length variable. To deal with these unique Python features, tremendous efforts may be required to design Python-to-SMT rules to support them.  
These unique difficulties result in Python not having a widely adopted symbolic execution engine. Most existing symbolic execution engines only support simple data types, such as integers and floats.
Some developers tried to overcome these obstacles with concolic execution \cite{sen2007concolic}, which combines symbolic execution with concrete execution on real values. However, concolic execution is a dynamic method that requires a runnable code environment. However, in many practical applications, such a code environment is often difficult to configure.


To address the above difficulties, we took the first step to perform a large language model-augmented symbolic execution engine, \method, for Python. \method is an LLM agent that can automatically call the Z3 \cite{de2008z3} SMT solver to solve path constraints in symbolic execution. On the basis of our LLMSym lies an introductory-level symbolic execution engine, The Fuzzing Book \cite{zeller2019fuzzing}, which only supports variables with explicit type hints in int/float types. The basis engine consists of two components. The first one is a control flow graph (CFG) explorer for extracting execution paths. The second one is a program for translating the collected execution path constraints into Z3 code using a pre-defined set of rules.
In our approach, we only adopt the CFG explorer in the engine for execution path collection. Instead of writing a program to translate Python constraints to Z3, we leverage an LLM to generate code for calling the Z3 solver to solve path constraints. This gives us more flexibility in solving some constraints that cannot be easily translated into Z3. During the generation of Z3 code, we propose a workflow with multiple-step generation and self-refinement to improve the correctness of the generated code. 

In general, \method solves a CFG path constraint with the following steps: (1) Given a program under test, \method first performs type inference to predict the type of variables inside the program. (2) For the execution path, we decompose it into chunks and generate the Z3 code with multiple steps. Before each generation step, \method queries a path-to-Z3 knowledge base to acquire code generation templates. We manually build this knowledge base to enable the generation of Z3 code to solve constraints on Python lists. (3) After generating code for a path chunk, we execute the generated code to check whether it contains bugs. If the generated code is buggy, \method tries to fix the code given the error message. If all fixes are unsuccessful, we regard this path as unsolvable and directly use an LLM solver to solve the constraints. (4) After \method finishes the code generation for an execution path, the Z3 solver will try to solve the generated constraints, and \method generates a final test case based on the solving results. All the code generation steps in \method are fulfilled with LLMs, which minimizes the effort of writing complicated logic for Z3 code generation.

To evaluate the LLMSym framework, we conduct experiments on a set of LeetCode problem solutions for unit test case generation. LeetCode programs contain plenty of control flows and list operations, which make them suitable for evaluating our symbolic execution engine. We build a dataset of 111 execution traces from 50 programs, and ask \method to solve these traces and generate test cases that can satisfy them. In the experiments, most components in \method is based on an efficient LLM GPT-4o-mini. In our experiments, we aim to answer the following research questions:

\begin{itemize}
    \item \textbf{RQ1}: Can \method generates correct Z3Py code to solve Python path constraints?

    \item \textbf{RQ2}: How do each module and different settings of the Z3 code generator work when solving a path constraint?

    \item \textbf{RQ3}: What are the strengths/weaknesses of the Z3 code generator and the LLM solver in solving path constraints?

    \item \textbf{RQ4}: How is the time and money cost of \method?

\end{itemize}

 The results show that \method can correctly solve around 60\% of the given execution traces and outperforms pure LLM-base approaches, proving the potential of using LLM to further guide symbolic execution by generating code that calls SMT solvers. Moreover, our approach is also cost-effective: as an LLM agent, its cost is less than one-third of a stronger model, GPT-4o, while achieving similar results.

To summarize, the main contributions of our paper are:

\begin{itemize}
    \item We propose \method, the first LLM agent-based system that augments an existing symbolic execution engine to solve more complex constraints. In \method, we put forward the idea of using LLM to call the Z3 solver for solving constraints in symbolic execution. 
    This allows us to build \method on a simple introductory-level symbolic execution engine without the laborious effort of writing lengthy code to translate path constraints into Z3.

    \item As far as we know, \method is the first Python symbolic execution engine that supports Python `list' and dynamic data types without concolic execution.
    It is also the first attempt to integrate LLMs with SMT solvers for test case generation.

    \item We create a dataset from LeetCode programs execution paths and evaluate \method using this dataset. We find that \method is capable of generating correct Z3 code and test cases for long execution paths. 
\end{itemize}

The remainder of this paper is organized as follows. Section 2 describes a motivating example. Section 3 describe the methodology of \method in details. Section 4 presents the experimental evaluation and answer to the RQs. Section 5 is threats to validity. The related works in introduced in Section 6, and Section 7 concludes our work.

\section{Motivating Example}

\begin{figure}[h]
  \centering
  \includegraphics[width=\linewidth]{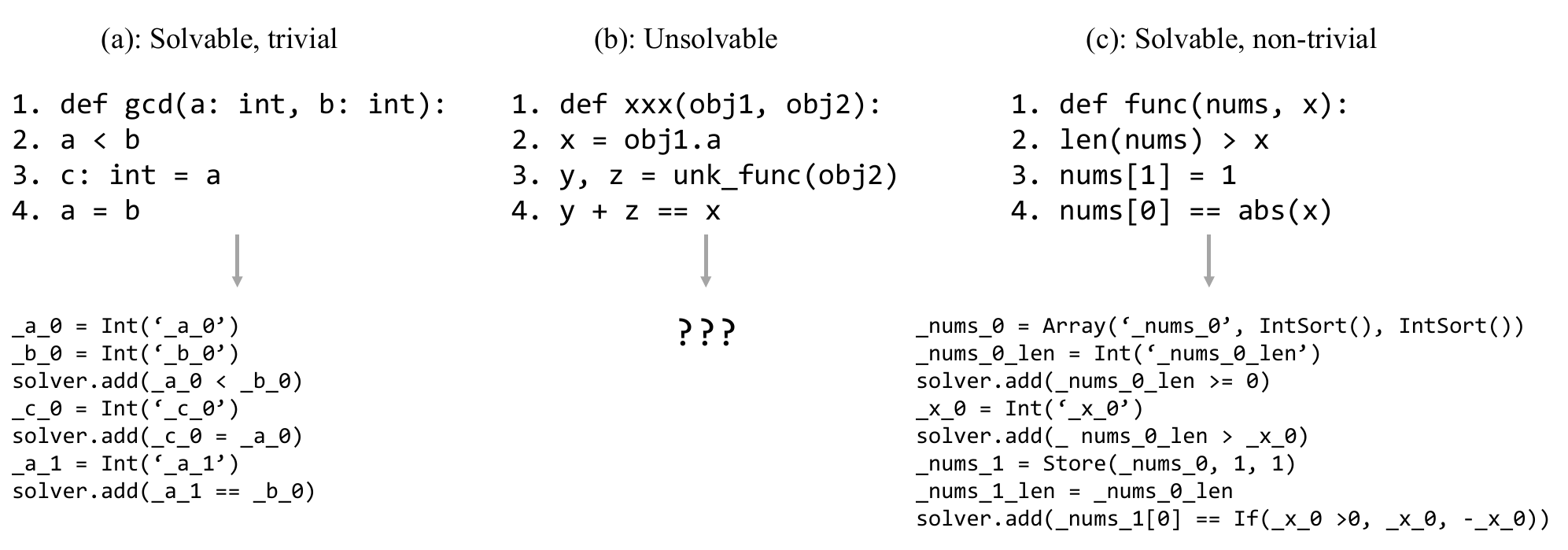}
  \caption{A motivating example of using LLMs to generate Z3 code for symbolic execution. LLMs are capable of generating complex code with non-trivial patterns (c).}
  \label{fig:motivation}
  \Description{The motivating example. (a): Trivial execution paths can be solved with Z3 in simple code. (b): A complicated path that is unsolvable by Z3. (c) A non-trivial path that is solvable by Z3, but requires extensive effort.}
\end{figure}

Figure~\ref{fig:motivation} shows a motivating example for LLM-empowered symbolic execution. Figure~\ref{fig:motivation} (a) (top) depicts a simple Python execution path with function definition, comparison operators, and variable assignments. In this execution path, all variables are explicitly annotated with a basic data type (int), and all operators are simple computational operators. In this case, we can easily generate the code which invokes Z3 to solve this path (Figure~\ref{fig:motivation} (a) (bottom)). In this example, we follow the format defined in \cite{zeller2019fuzzing}: variables are transformed into the single static assignment (SSA) form to keep variable values before assignments. We can see that the translation from the execution path to Z3 is simple: we convert variables to SSA and then add every statement to the solver. This can be implemented without extensive coding effort: The Fuzzing Book \cite{zeller2019fuzzing} used 76 lines of code to translate one line of execution path into Z3 code. It first uses abstract syntax tree (AST) rewriting to generate Python statements in SSA form, then translates the SSA statements into Z3-compatible form with simple rules.

However, traditional rule-based symbolic execution has its limitations. The major limitations include difficulties in handling complex data structures and external function calls. For instance, in Figure~\ref{fig:motivation} (b) (top), there are user-defined data types and functions with external functions that cannot be interpreted as SMT-compatible constraints. As a result, it becomes infeasible to generate a Z3 code snippet to accurately describe the constraints for this execution path.
Traditionally, researchers tend to overcome these situations by concolic execution \cite{sen2007concolic}, where test inputs are directly executed with concrete values. However, concolic execution also has its limitations. First, relying on concrete inputs might cause concolic execution to have difficulty reaching certain paths. Second, directly executing unverified test inputs may introduce undesired security risks \cite{li2024holistic}.  
To address these complex, non-linear constraints, a natural approach is to leverage large language models (LLMs) for solving. LLMs have demonstrated considerable capability in logical reasoning, including tasks such as solving mathematical problems \cite{cobbe2021training} or logic puzzles \cite{srivastava2022beyond}. Similarly, we can harness the reasoning ability of LLMs to solve path constraints in symbolic execution.

In addition to the two types of path constraints above, there is another type that lies in between: the constraints are theoretically solvable by a solver, but require non-trivial reasoning by domain experts. Figure~\ref{fig:motivation} (c) (top) demonstrates an example execution path that contains list indexing, list assignments, and a simple built-in function (abs()). For these operations, designing rules to map them to Z3 expressions can be difficult and time-consuming. However, human developers with knowledge of Z3 can still write a program to solve these constraints in only a few lines of code (Figure~\ref{fig:motivation} (c) (bottom)). Based on this observation, we propose utilizing LLMs to automatically generate Z3 code for such non-trivial constraints, which is one of the key contributions of \method.

Previous works have shown that LLMs are capable of calling the Z3 solver \cite{ye2024satlm, pan2023logic, wang2024dataflow}, but they mainly focus on solving simple logical/arithmetic problems with only a few lines of code. On the other hand, symbolic execution brings new challenges to Z3 code generation as it requires us to generate longer, more complicated Z3 code to solve lengthy execution paths with diverse data types and execution behaviors.

\section{Approach}

\begin{figure}[h]
  \centering
  \includegraphics[width=\linewidth]{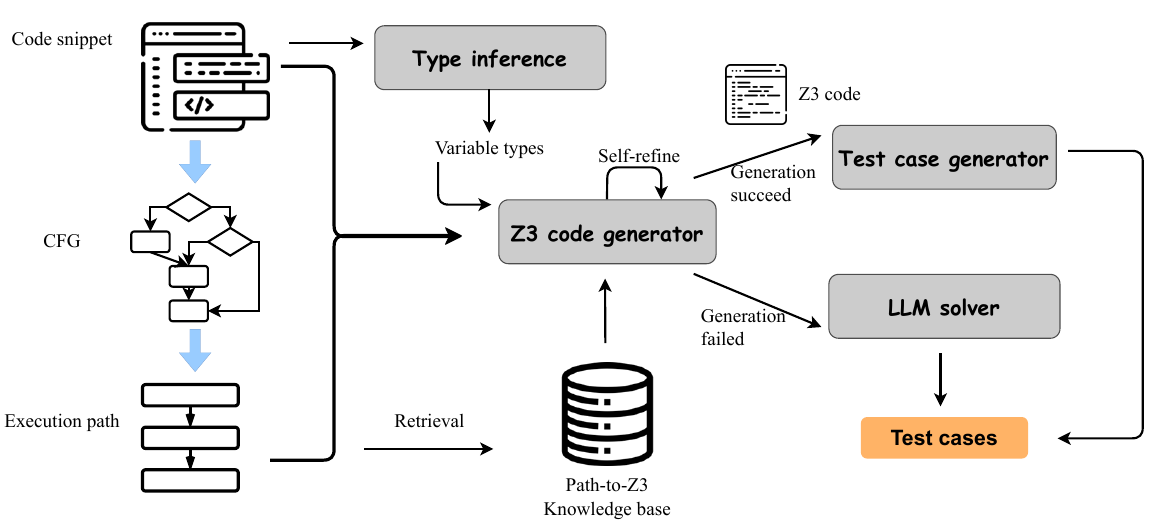}
  \caption{The overview of LLMSym workflow.}
  \label{fig:overview}
  \Description{}
\end{figure}

Figure~\ref{fig:overview} depicts the overall workflow of our \method. \method is a system with two components: an execution path extractor, and an LLM agent which generates test cases based on execution paths. As demonstrated, for a given Python program under test, we first generate its control flow graph (CFG), then traverse the CFG to extract execution paths. After we extract an execution path, an LLM-powered Z3 code generator will translate this path into Z3 constraints. To enable the correct generation of Z3 code, we build a knowledge base with path-to-Z3 pairs and retrieve templates from this knowledge base to guide the code generator. If the initially generated Z3 code is buggy, the code generator will try to fix it using a self-refine mechanism. If the Z3 code for the complete execution path is generated successfully, then a test case generator LLM takes the solver's results as input and interprets it as a complete test case. If the Z3 code generation is unsuccessful, we directly use LLM to solve the path constraints and produce test cases. The main components of the LLMSys agent are based on a sub-optimal and cost-efficient LLM GPT-4o-mini, which helps in reducing the overall cost.

\subsection{Execution path extractor}
The main body of the path extractor is built on the implementation in The Fuzzing Book \footnote{The code for path extraction can be found at \url{https://www.fuzzingbook.org/html/SymbolicFuzzer.html}}\cite{zeller2019fuzzing}. Basically, it first builds the program CFG from its abstract syntax tree (AST), then uses breadth-first search to extract all paths from the CFG. In order to successfully create execution paths that LLMs can comprehend, we made the following improvements to the original implementation:

\begin{itemize}
    \item In the CFG parsing module, we add support to class definitions so that We can build CFGs for Python functions that are located inside a class.

    \item For `for' loop statements, we add a number to the execution path node to denote which iteration of the loop the path has reached.

    \item In the CFG, each node corresponds to a Python statement. We categorize all statements into 3 types. The first type is `enter', where we encounter a function definition statement, and enter this function during execution. The second type is `condition', where we encounter a branch/loop condition in a `if'/`while'/`for' statement. The third type is `expression': all other CFG nodes belong to this type, most of which are assignment statements.
\end{itemize}

\begin{figure}[h]
  \centering
  \includegraphics[width=\linewidth]{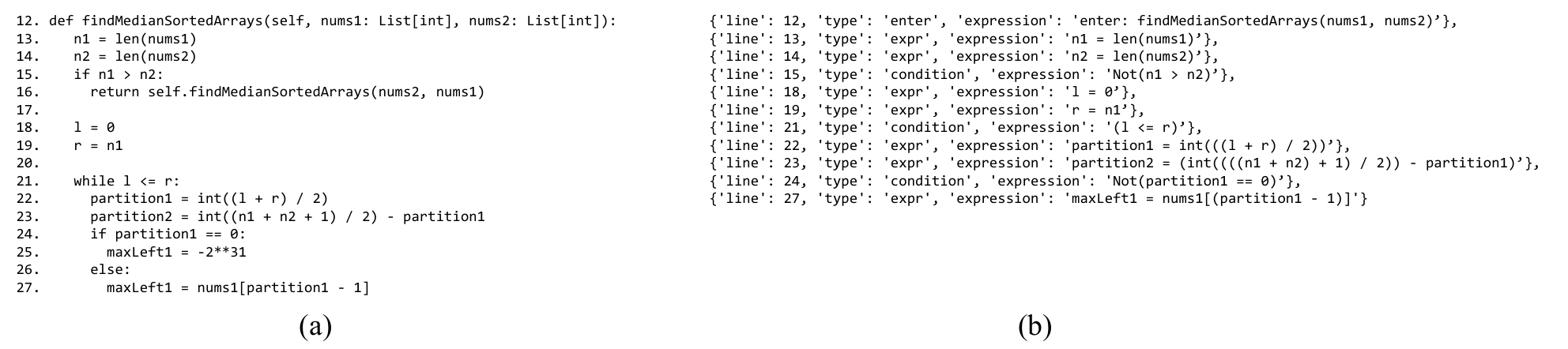}
  \caption{An example of an extracted execution path (b) for the program (a).}
  \label{fig:path}
  \Description{}
\end{figure}

An example of an extracted execution path is shown in Figure~\ref{fig:path}. Every execution step consists of a line number, a statement type, and the statement itself. After the execution path is extracted, we split it into chunks before providing it to the Z3 code generator. This is because we find that the Z3 code generator is more likely to err when the input execution path sequence is long. In our default setting, we split the path by lines: each chunk only contains one statement in the execution path. We also adopt an alternative approach that splits the path by condition statement, and we will discuss this further in the follow-up sections.

We need to be aware that our improvements do not mitigate the path extractor from its intrinsic weakness: the path explosion problem, as optimizing path extraction is not the main goal of our work.

\subsection{LLM agent for constraint solving}

\subsubsection{Type predictor}
\begin{figure}[h]
  \centering
  \includegraphics[width=\linewidth]{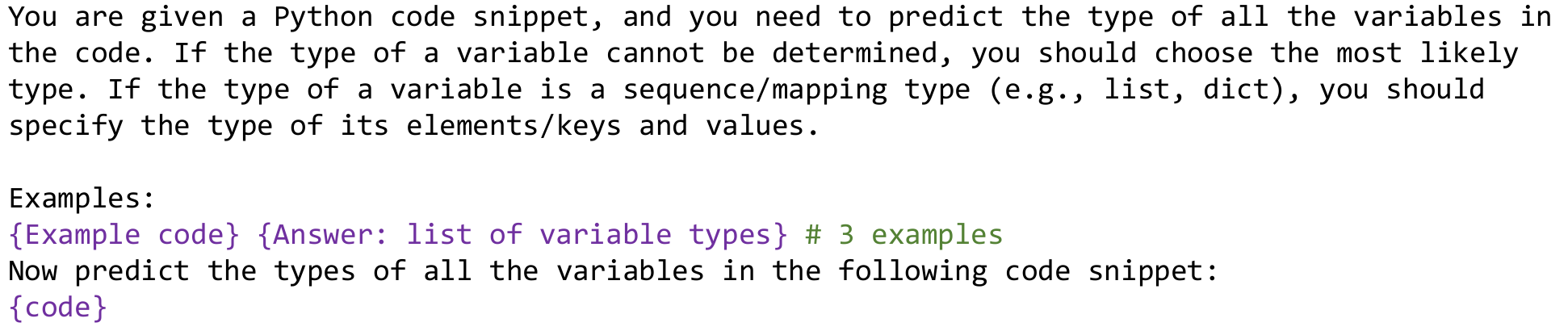}
  \caption{The prompt template for the type predictor.}
  \label{fig:type}
  \Description{}
\end{figure}

One of the critical features of Python, different from many other program languages, is its dynamically typed variables. In Python, when we declare a new variable, we do not need to annotate its type explicitly. On the other hand, in Z3, symbolic variables must be assigned with a fixed type when initialized. In order to translate Python expressions to Z3 expressions, we first need to know the variable types in the execution path. So, before we start generating Z3 code, we use an LLM to predict the variable types in the program under test. Specifically, we ask the LLM to gather all variables and input arguments in the program, and predict their types. We use few-shot examples to provide relevant knowledge for type inference, and regulate the output format of the type predictor. Here, we made an approximate assumption: the type of a variable does not change during the execution, which can be applied to the majority of variables. Figure~\ref{fig:type} shows the prompt template for our type predictor.

\subsubsection{Z3 code generator}
The core of our LLM agent in \method is to translate extracted path constraints into Z3 code. For a given execution path, the Z3 code generator generates the Z3 constraints for the path in multiple steps. In each step, the code generator takes three pieces of information as input: an execution path chunk (ends with a condition statement), the variable types (generated by the type predictor), and the previously generated Z3 code (generated by the Z3 code generator in previous steps).

To translate a Python expression in the execution into the Z3 Python API (Z3Py) \footnote{https://z3prover.github.io/api/html/namespacez3py.html}, the Z3 code generator is required to perform two steps of reasoning: first, similar to \cite{zeller2019fuzzing}, all variables should be transformed into SSA. This is not a difficult step since the knowledge of SSA is already processed by LLMs. We designed a few examples to demonstrate the variable naming specifications (e.g., the 2nd re-assignment of variable `x' should be written as `\_x\_2' in our SSA setting). 
To further enhance the memory of SSA for the LLM, we introduce a variable environment to store the SSA indices for each variable. In each generation with the Z3 code generator, we give the current SSA variable environment to the LLM, and ask it to return the updated variable environment along with the generated Z3Py code.

\begin{figure}[h]
  \centering
  \includegraphics[width=0.7\linewidth]{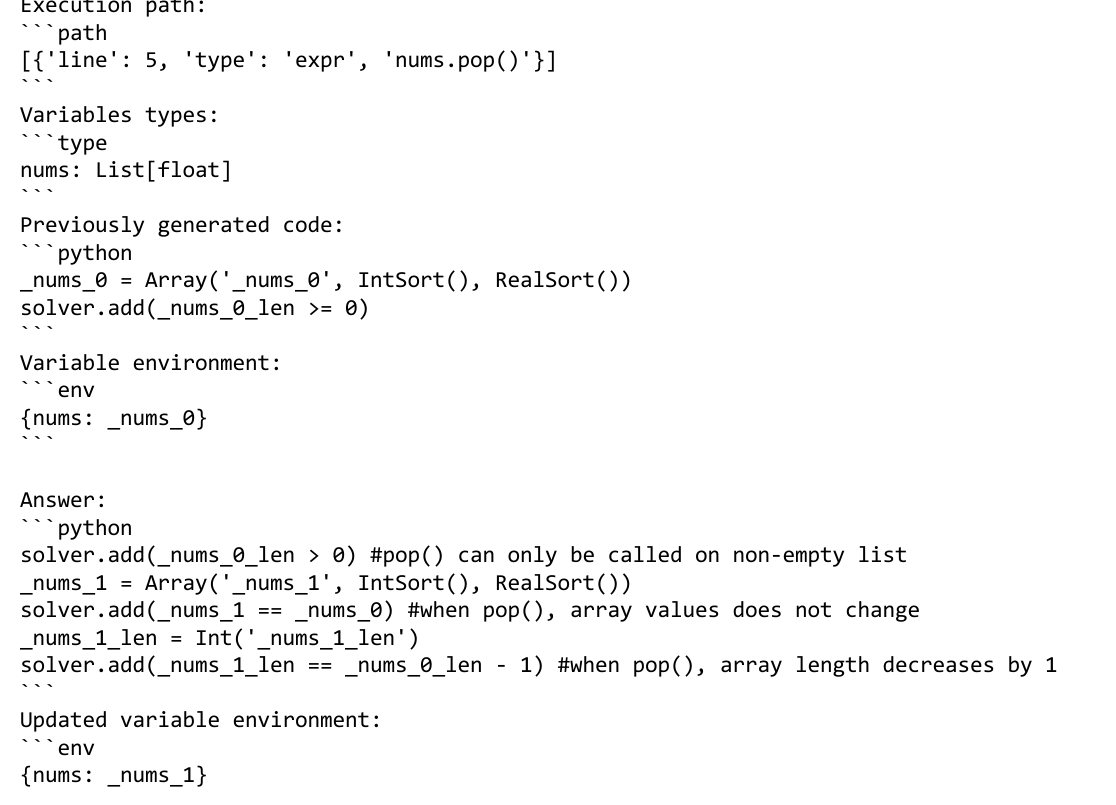}
  \caption{An example for the path-to-Z3 template in the retrieval knowledge base.}
  \label{fig:retrieve}
  \Description{}
\end{figure}

\begin{figure}[h]
  \centering
  \includegraphics[width=\linewidth]{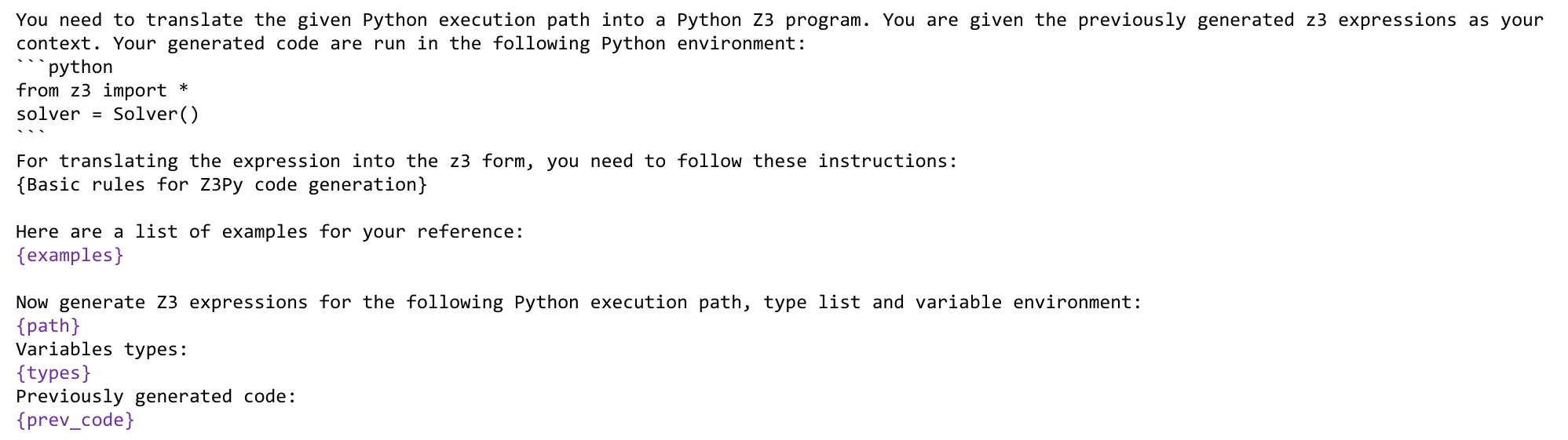}
  \caption{The prompt template for the Z3 code generator.}
  \label{fig:gen}
  \Description{}
\end{figure}

A more difficult reasoning step is to accurately translate the constraints semantics in the Python statement into the Z3Py form. During practice, we find that LLMs often make syntax errors when generating Z3Py code, which is understandable since Z3 code rarely occurs in the training corpus of LLMs. Moreover, when we encounter complex data structures (e.g., Python `list'), it is non-trivial to translate its operations and constraints into Z3Py. To improve the syntactical and semantical correctness of generated Z3 code, we leverage retrieval augmented generation techniques to provide expert knowledge to the code generator LLM. Specifically, we build a Python-to-Z3 knowledge base with execution paths and Z3Py code templates. In each template, we implement a Python operation that cannot be directly interpreted to Z3Py by LLM. One of the main targets of the knowledge base is to support Python operations on `list'. In \method, we approximate Python lists using `Z3.Array', because the `Array' data structure has an infinite length, which is suitable for Python lists which has an unfixed length. To represent the length of a Python list, we additionally define a `Z3.Int' symbolic length variable for each list so that the Z3 solver can reason about Python list lengths. Currently, we support 4 types of list operations in our knowledge base: list initialization, list indexing (read list values), list assignments (write list values), and list append. Table~\ref{tab:list} shows how we implemented these functionalities in Z3Py. Apart from list operations, we also include several other operations in our knowledge base, including general assignments, numerical computations, etc. In total, we design 14 Python-to-Z3 templates in the knowledge base for retrieving relevant information. Each template contains a full input-respond example of the Z3 code generator, which includes an execution path chunk (with one or several statements), an input SSA variable environment, with the target Z3Py code, and an updated variable environment. A complete example of a retrieval template is demonstrated in Figure~\ref{fig:retrieve}.
When the Z3 code generator tries to generate Z3 code for an execution path chunk, it first searches the Python-to-Z3 knowledge base, using the path chunk as the query. In the knowledge base, the execution path for each template is used as the key. All keys and queries are encoded to vectors using the Sentence-BERT \cite{reimers-2019-sentence-bert} model, and we use the L2-norm as the distance metric for retrieval. For each retrieval query, we return the most similar templates, and these templates are used as the few-shot examples for Z3Py code generation. The complete prompt template for the Z3 code generator is shown in Figure~\ref{fig:gen}.

\begin{table}
  \caption{Examples of translating Python list operations into Z3Py constraints for \method.}
  \label{tab:list}

  \begin{tabular}{ccc}
        \toprule
        Operation & Python execution path & Z3Py code example \\
        
        \midrule
        List initialize & enter func(n1: List[int]): & \makecell[l]{\_n1\_0 = Array('\_n1\_0', IntSort(), IntSort()) \\ \_n1\_0\_len=Int('\_n1\_0\_len') \\ solver.add(\_n1\_0\_len >= 0)} \\
        \midrule

        List length & len(n)>5 & \makecell[l]{solver.add(\_n\_0\_len) > 5} \\
        \midrule

        List indexing & lst[i] == j & \makecell[l]{solver.add(\_lst\_1[\_i\_0] == \_j\_0)}\\
        \midrule

        List assignment & lst[i] = 2 & \makecell[l]{\_lst\_2 = Store(\_lst\_1, \_i\_0, 2) \\ \_lst\_2\_len = \_lst\_1\_len \# list length is not changed} \\
        \midrule

        List append & n.append(x) & \makecell[l]{\_n\_1 = Store(\_n\_0, \_n\_0\_len, \_x\_0) \# n.append(x) \\ \_n\_1\_len = \_n\_0\_len + 1 \#increase length by 1} \\
        \midrule

        List pop & n.pop() & \makecell[l]{solver.add(\_n\_0\_len > 0) \#before pop(), check length \\ \_n\_1 = \_n\_0 \#Array values do not change \\ \_n\_1\_len = \_n\_0\_len - 1 \#decrease length by 1}\\
        \midrule

        List negative index & lst[-2] == z & \makecell[l]{ solver.add(\_lst\_0[\_lst\_0\_len - 2] == \_z\_2) \\
        solver.add(-2 >= -\_lst\_0\_len)}\\
        \bottomrule
    \end{tabular}
\end{table}

Even if we integrate multiple-step generation and retrieval templates, the Z3 code generator may still generate buggy code. When the Z3Py code for a chunk is generated, we first concatenate it with the previously generated code and then execute it. If the execution fails with an error, the code generator tries to fix it with a self-relection mechanism. We input the buggy code and the error message into an LLM, and ask it to fix the code. In our experiments, we keep tying this self-repair process 3 times for each chunk. If all fixing attempts fail, we terminate the Z3 code generation and directly ask an LLM (the LLM solver in Figure~\ref{fig:overview}) to solve the execution path constraints.

\subsubsection{Test case generator}
After the Z3 code generator successfully generates the Z3Py code for a complete path, the test case generator module will generate the final test case based on the solving results of the generated Z3 constraints. The Z3 solver will output the values of all solved SSA variables, and the test case generator needs to analyze these values, find variables that are related to the test input, and build a test case from the value of these variables.

\begin{figure}[h]
  \centering
  \includegraphics[width=\linewidth]{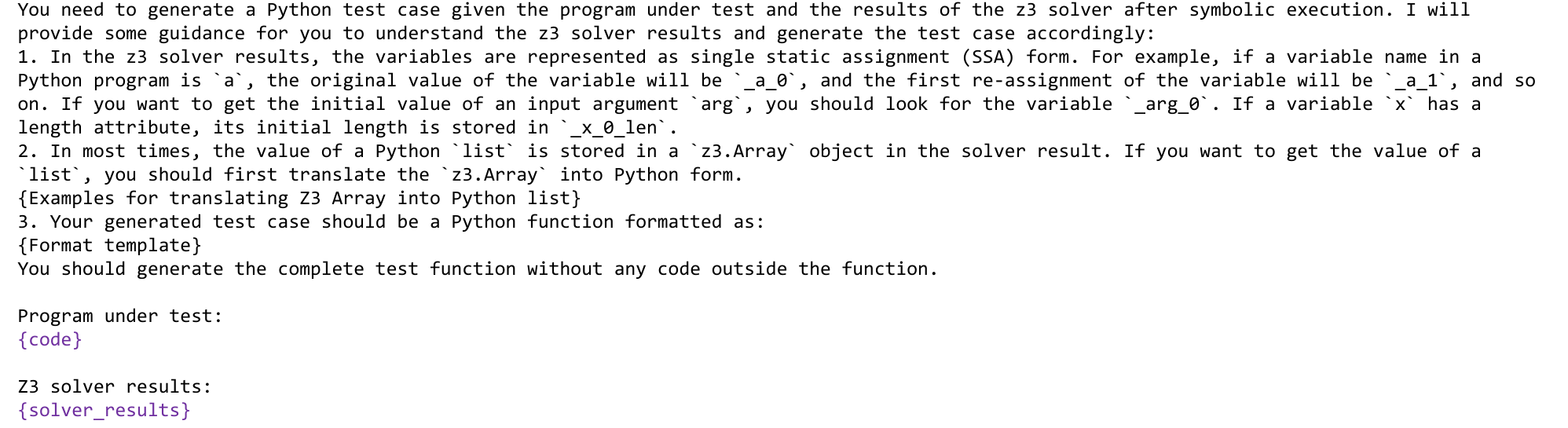}
  \caption{The prompt template for the test case generator.}
  \label{fig:test}
  \Description{}
\end{figure}

Among all types of variables returned by the Z3 solver, the `Z3.Array' type needs special consideration. The value of Z3 Array variables cannot be easily interpreted as a Python list: it is represented by first defining a constant array (e.g., K(Int, 1)), then applying `Store' operations (e.g., Store(x, 0, 1) means set x[0] to 1) on it. For example, an Array `Store(K(Int, 0), 2, 3)' can be transferred to a Python list as [0, 0, 3, 0, ...]. In \method, with the Array value and the length value, the test case generator should be able to restore them as a Python list in a test input. For this module, we choose GPT-4o as the backend model because it shows significantly better performances in interpreting Z3 Arrays than GPT-4o-mini. Figure~\ref{fig:test} demonstrates the prompt template used in the test case generator.

\subsubsection{LLM solver}
\begin{figure}[h]
  \centering
  \includegraphics[width=\linewidth]{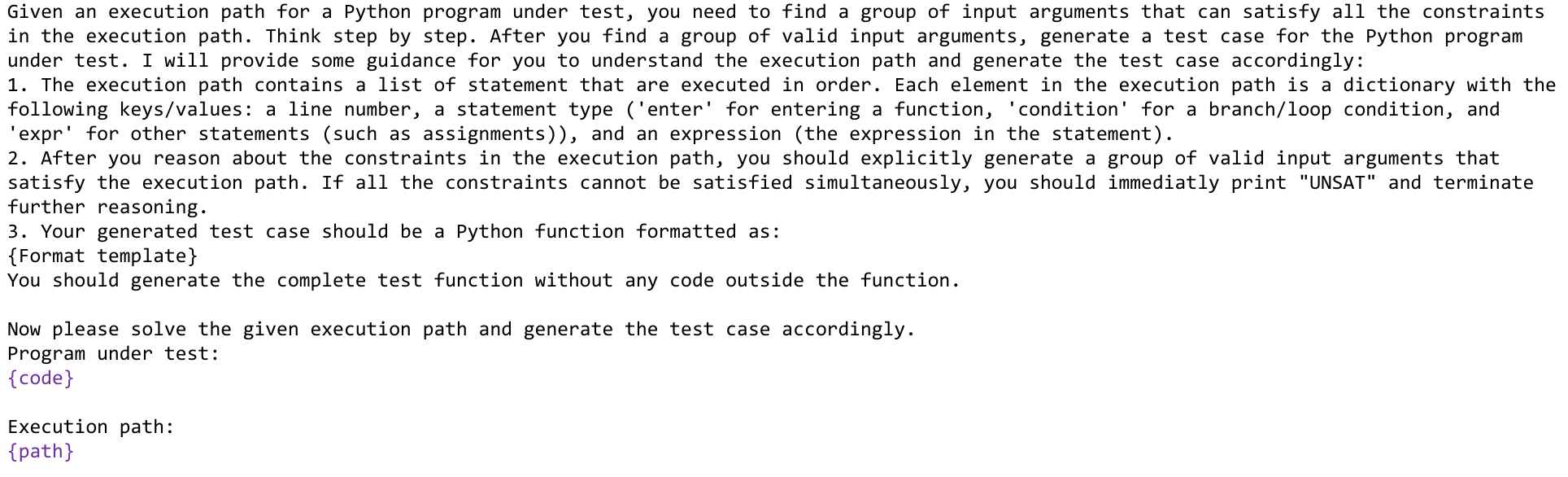}
  \caption{The prompt template for the LLM solver.}
  \label{fig:solve}
  \Description{}
\end{figure}

When the Z3 code generator is incapable of generating Z3 code to solve path constraints, we directly seek the LLM to accomplish this task. In the LLM solver, the LLM is given the program under test and the execution path. The LLM solver is asked to solve these constraints directly, produce possible values for input arguments, and then generate a test case from the produced values. 
Similar to the Z3 code generator, when the LLM solver is asked to solve the constraints of an execution path, it also needs to determine whether this path can be satisfied. If the LLM solver predicts that a path cannot be satisfied, it will terminate generating the answer, allows us to explore other paths.
As this module is not the key contribution of our \method, we only design a simple prompting mechanism in our current implementation and leave the improvement of the LLM solver as future works. Figure~\ref{fig:solve} shows the prompt template of the LLM solver.

\section{Experiments}

\subsection{Experiment settings}
In \method, apart from the test case generator module, all other modules are powered by the cost-efficient LLM GPT-4o-mini. The test case generator is built upon GPT-4o. We implement the \method agent using the AutoGen \cite{wu2024autogen} library.

\subsection{Dataset}

To evaluate the capability of \method, we build a small-scale dataset from the online programming platform LeetCode. We select 50 problems from LeetCode, whose Python solutions \footnote{The Python solutions are collected from a public GitHub repository \url{https://github.com/walkccc/LeetCode}.} with up to 25 lines of code. These programs contain statements with multiple data types and control flows. For each problem, we run the example test cases in their problem description and collect the execution traces. This results in a total of 111 paths. We further truncate the traces with a max length of 20, and the average trace length in our dataset is 12.3.
The collected execution traces do not contain information on branch selection, so we use the CFG path extractor in \method to find the complete execution path corresponding to the execution trace. Our main task is to let \method solve the constraints for these execution paths and generate test cases. After we obtain the generated test cases, we execute them to find whether the generated test case can cover the same path as the example test case. Due to the lack of explicit variable type annotations and the presence of Python list data structures, our backbone symbolic execution engine \cite{zeller2019fuzzing} can solve \textbf{None} of these traces.


\begin{table}
  \caption{The number (rate) of passed paths solved by \method in our LeetCode dataset.}
  \label{tab:res}

  \begin{tabular}{ccc}
        \toprule
        SAT & Execution pass & Path correct \\
        \midrule
        99 (89.2\%) & 97 (87.4\%) & 70 (63.1\%) \\
        \bottomrule
    \end{tabular}
\end{table}

\subsection{RQ1: Capability of \method}
Table~\ref{tab:res} shows the results of \method on our LeetCode dataset. We measure the number of generated results (test case or ``UNSAT'') that can solve path constraints (``SAT"), produce runnable test cases, and correctly fulfill the given execution trace. From all 111 results generated from \method, over 63\% percent of them can produce the exact same execution trace as the given input. This shows the potential of using LLMs to generate code for solving symbolic path constraints.

\begin{figure}[h]
  \centering
  \includegraphics[width=\linewidth]{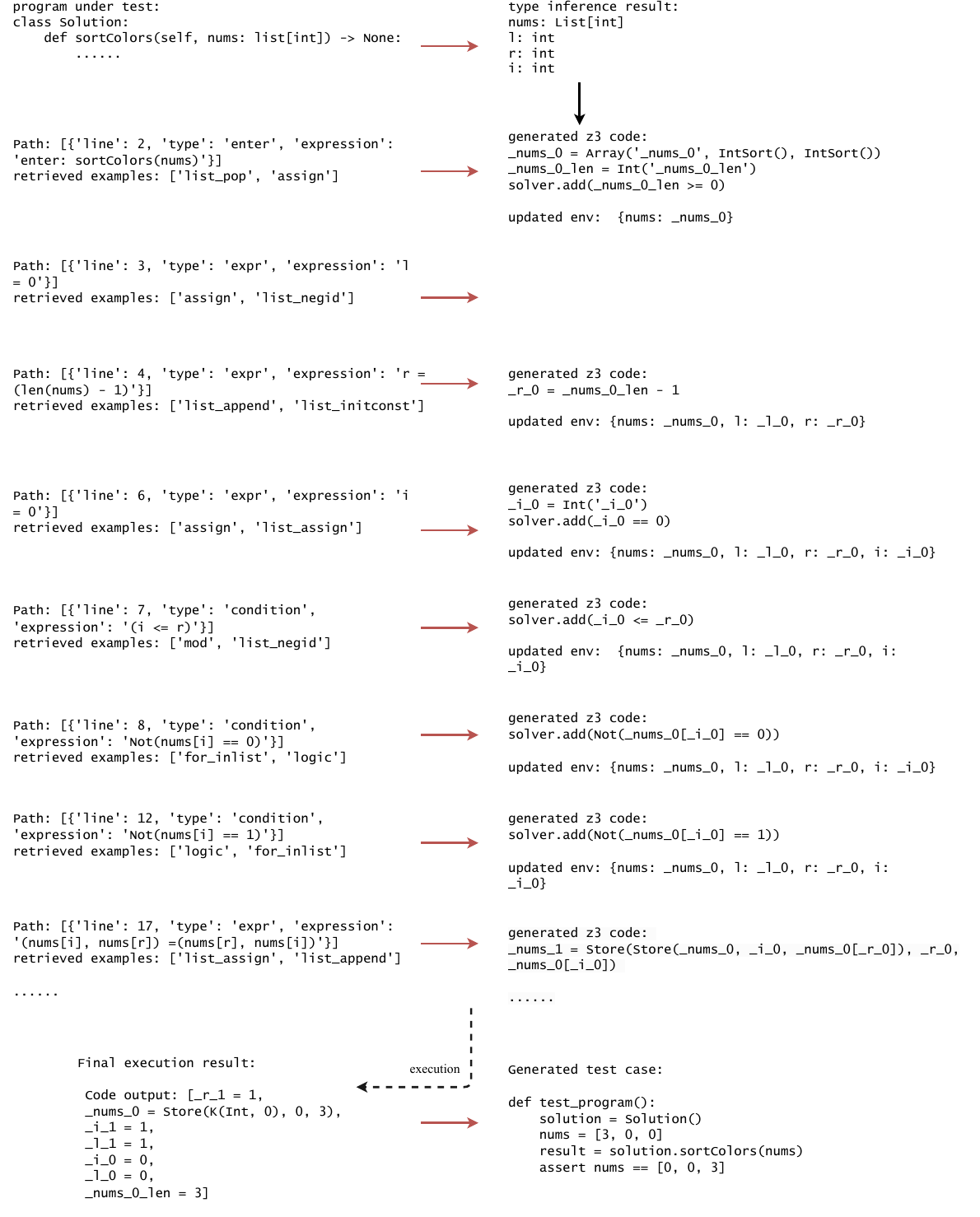}
  \caption{An example of the Z3Py code and test case correctly generated by \method. Due to the path length, we only display a part of it.}
  \label{fig:eg1}
  \Description{}
\end{figure}

Figure~\ref{fig:eg1} shows an example of a success generation by \method. We can see that during the generation process, the type predictor first correctly generates variable types for all variables in the program. In generating Z3Py code for each Python statement, the retriever first retrieves relevant templates from the knowledge base, which is extremely important for non-trivial operations. For example, in line 12, the retriever successfully returned the ``list assignment'' template that helped generate the correct Z3Py statement. Meanwhile, for simple operations, the retrieval results are not so important. For example, in lines 7 and 8, the retriever returns irrelevant results, but the code generator can still generate the correct Z3Py code. An interesting finding emerges from line 12: the LLM code generator correctly generates the Z3Py code for an advanced list operation. It implements ``swap list elements" with two Array assignment operations, while similar patterns do not exist in our template knowledge base. This suggests that the LLM code generator can understand the semantics of path-to-Z3 templates (in this case, the ``list assignment'' template) and flexibly compose the basic operations in these templates to implement more complex operations. From this example, we can see that the LLM code generator shows strong capability in generating Z3Py code with the retrieval and memory mechanism designed by ours. 

\begin{figure}[h]
  \centering
  \includegraphics[width=\linewidth]{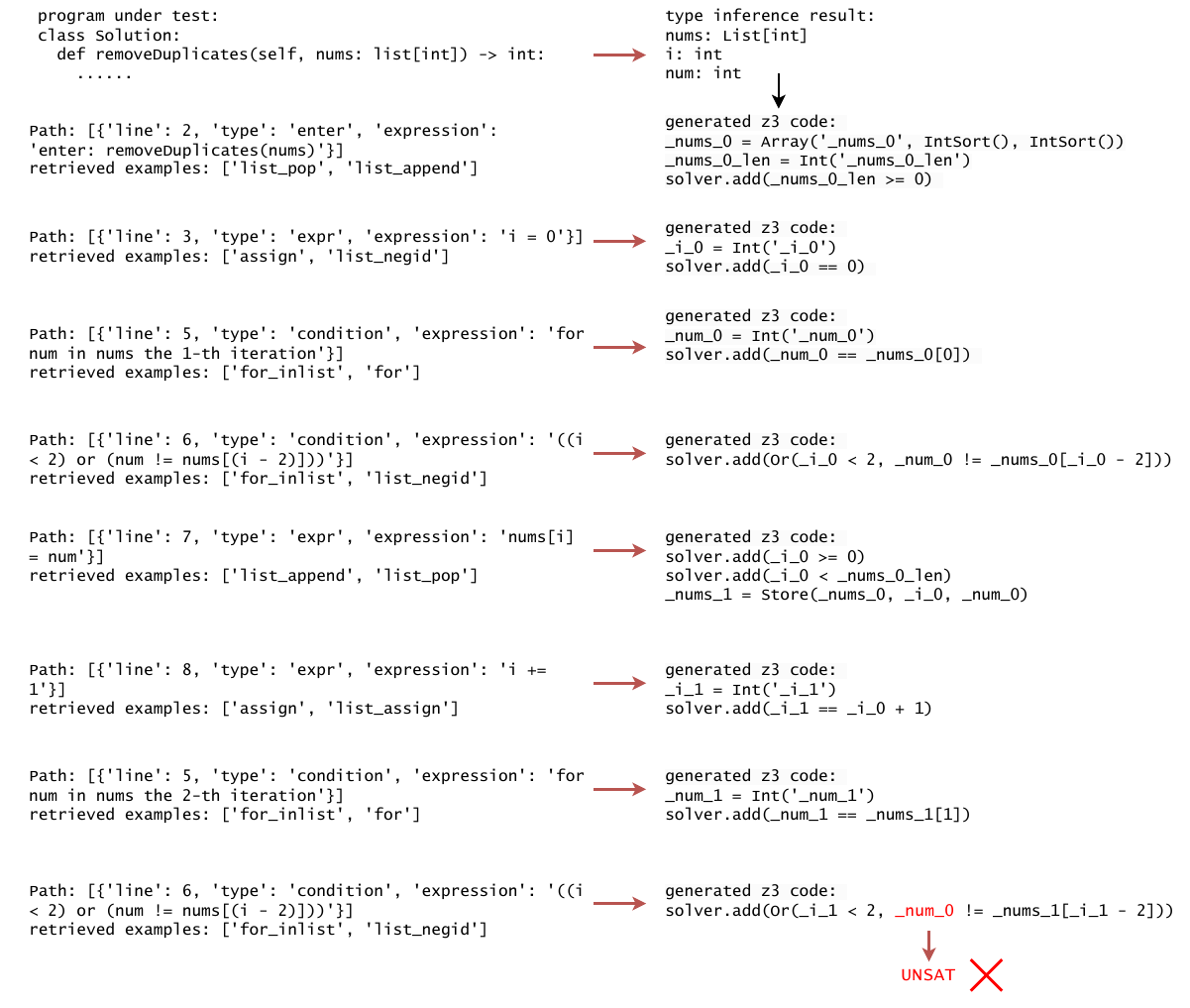}
  \caption{An example of an incorrect generation of \method.}
  \label{fig:fail}
  \Description{}
\end{figure}

Figure~\ref{fig:fail} shows a failed example of \method. This failed generation finally lead to the generate constraints cannot be satisfied. In this example, the Z3 code generator generates correct Z3Py code until it reaches the 8th statement in the execution path: ``(i
< 2) or (num != nums[(i - 2)])''. Before this statement, the Python variable ``num'' corresppond to the Z3Py SSA variable ``\_num\_1'', and this statement do not make new assignments to ``num'', so the Z3 code generator should use variable ``\_num\_1'' in its generated code. However, the LLM failed to generate the correct SSA variable reference, and use ``\_num\_0'' instead. This will cause the generated Z3 constraints differ from the given path constraints, and eventually, in this case, fail to solve the given path when it is actually satisfiable.

\begin{tcolorbox}[size=title,breakable]
\textbf{Answer to RQ1:} \textcolor{black}{\method can generate correct Z3Py code to solve path constraints for symbolic execution. For non-trivial list operations, our retrieval mechanism allows the LLM to acquire knowledge on translating Python constraints to Z3 and generate the correct Z3Py code thereafter. However, sometimes the Z3 code generator still make mistakes, such as generating incorrect SSA variables.}
\end{tcolorbox}

\subsection{RQ2: Submodule analysis}
To study the effectiveness of each module in \method, we first investigate the impact of different settings of the retriever. Table~\ref{tab:res-retrieve} shows the results of \method on our LeetCode dataset under different number of retrieved templates. We find that when the retriever returns two templates, \method has the highest pass rate for execution paths, and increasing the number of retrieved samples will cause a slight drop in the passed paths. We believe that only returning one template may miss the key relevant template, while returning too many templates may mislead the LLM into generating incorrect code.

\begin{table}
  \caption{The number of passed paths solved by \method with different numbers of templates returned by the retriever.}
  \label{tab:res-retrieve}

  \begin{tabular}{cccc}
        \toprule
        No. of templates & SAT & Execution pass & Path correct \\
        \midrule
        1 & 91 & 87 & 54 \\
        2 & 97 & 94 & \textbf{65} \\
        3 & 100 & 100 & 62 \\
        4 & 90 & 90 & 61 \\
        5 & 95 & 93 & 62 \\
        \bottomrule
    \end{tabular}
\end{table}

We further study the efficacy of the retriever by measuring their recall@k in retrieving relevant templates. The results are shown in Table~\ref{tab:res-recall}. In this table, we exclude path constraint statements with only a simple condition (because the LLM can correctly translate them to Z3Py without external templates). On average, for over half of statements in our dataset, the retriever can return the relevant template as the most correlated retrieval result. This indicates the strong capability of the retriever. Nevertheless, there is still room for improvement in the retrieving module. Even if we return 5 retrieval templates, there are still 19\% of the path constraint statements that do not successfully retrieve a relevant template.

\begin{table}
  \caption{The recall@k of the \method retriever in retrieving relevant templates.}
  \label{tab:res-recall}

  \begin{tabular}{cccc}
        \toprule
        Recall@1 & Recall@2 & Recall@5 \\
        \midrule
        55.8\% & 64.2\% & 81.5\% \\
        \bottomrule
    \end{tabular}
\end{table}

\begin{figure}[h]
  \centering
  \includegraphics[width=\linewidth]{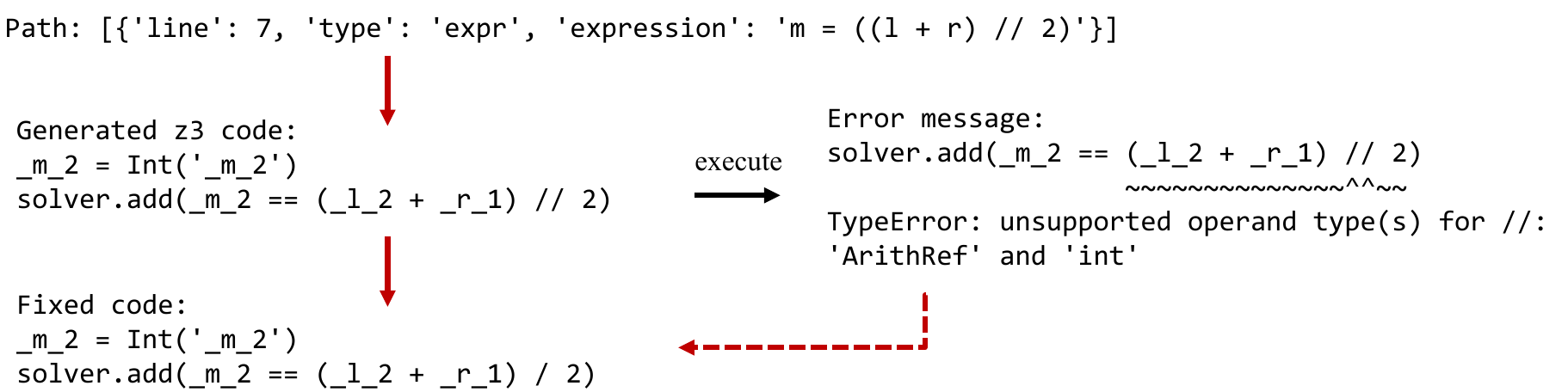}
  \caption{An example of the self-refine module successfully fixes a bug in generated Z3Py code.}
  \label{fig:eg-fix}
  \Description{}
\end{figure}

We perform a case study to demonstrate how the self-refine module fixes the generated code. Figure~\ref{fig:eg-fix} shows an example of a successful fix by our self-refine mechanism. For the input Python statement, the Z3 code generator first converts it into Z3Py without changing any operators. However, the Python floor division operator ``//'' is not supported by Z3, so when we execute the generated code, it returns a type error. The self-refine module receives the buggy code with the error message and successfully generates the fix. In the fixed code, the operator ``//'' is replaced by ``/'' because, in Z3Py, the ``/'' division operation on integer variables is naturally equivalent to the ``//'' operation in Python: it rounds the division result down to the nearest integer.

\begin{figure}[h]
  \centering
  \includegraphics[width=\linewidth]{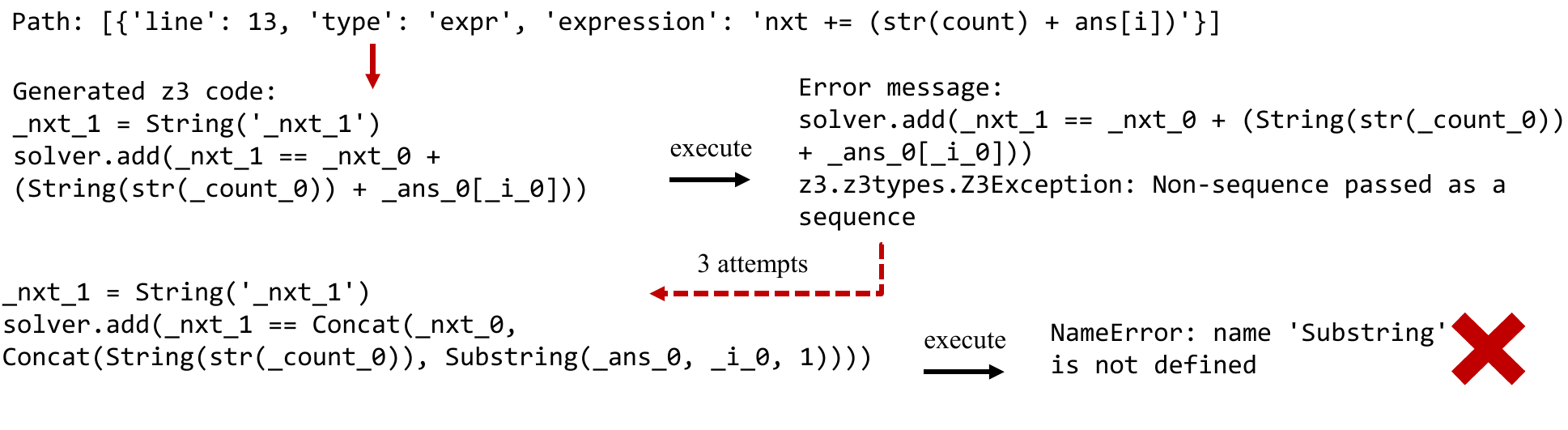}
  \caption{An example of the self-refine module fails in fixing a bug in generated Z3Py code.}
  \label{fig:eg-fixfail}
  \Description{}
\end{figure}

Figure ~\ref{fig:eg-fixfail} shows a failed example of the self-refine module in \method. The input Python code converts an integer to a Python string, although such operation is theoretically supported by Z3 \footnote{See \url{https://microsoft.github.io/z3guide/docs/theories/Strings/\#strfrom\_int-i-strto\_int-s---convert-to-and-from-non-negative-integers} for examples.}, its Z3Py API is rarely used in practice. From the example we can see that the LLM do not have the knowledge for converting an integer to a string, so it tries to generate different non-existing APIs in the three attempts, and all attempt are unsuccessful. We can conclude that when using the LLM to generate Z3Py code for non-trivial operations that are seldom seen in its training corpus, it is extremely important to provide high-quality examples to supplement it with new knowledge.

We further compare two path chunking strategies: chunk by line (our default setting) and chunk by condition statements (discussed in Section 3.1). Table~\ref{tab:res-chunk} shows the results of chunking by condition statements under different retrieval settings. We can observe that in all settings, chunking by condition results in fewer solved paths than chunking by line. This difference in performance is predictable because when we ask the LLM to generate Z3Py constraints for multiple Python statements, it is harder than generating for a single statement, and the LLM is more likely to make mistakes.

\begin{table}
  \caption{The number of passed paths solved by \method with path inputs split by condition statements (and the differences to the results of chunking by line).}
  \label{tab:res-chunk}

  \begin{tabular}{cccc}
        \toprule
        No. of templates & SAT & Execution pass & Path correct \\
        \midrule
        1 & 93 & 91 & 48(-7) \\
        2 & 83 & 80 & 49(-16) \\
        3 & 84 & 82 & 48(-14) \\
        4 & 82 & 79 & 45(-16) \\
        5 & 82 & 79 & 45(-17) \\
        \bottomrule
    \end{tabular}
\end{table}

\begin{tcolorbox}[size=title,breakable]
\textbf{Answer to RQ2:} \textcolor{black}{Both the retrieval and self-refine modules play a key role in improving the correctness of the Z3 code generator. For the retriever, different numbers of retrieved examples can affect the performance of the LLM code generator. Meanwhile, adding more templates in the retrieval knowledge base can allow the code generator to support more complex operations. Considering the results on recall@k, the retriever is generally powerful but can be further improved in the future. As for the self-refine module, it can fix some of the bugs in the generated Z3Py code, but if the input Python statement is beyond the ability of the code generator itself, then the self-refine module is unlikely to fix the buggy Z3Py code. For the path chunking strategies, split by lines achieves better results than split by condition statements.}
\end{tcolorbox}

\subsection{RQ3: comparison between Z3 code generator and LLM solver}
\method adopts two approaches in solving Python path constraints. The first one is using the Z3 code generator to generate Z3Py code for solving. The second one is to use an LLM for solving when the Z3 code generator fails. It is important to analyze the performances of both components and find whether the Z3 code generator can outperform the vanilla LLM solver with basic prompts.

\begin{table}
  \caption{The number and pass rates for the Z3 code generator (Z3) and the LLM solver (LLM) in solving path constraints. Notice that the sum of the solutions generated by the Z3 code generator and the LLM solver is not equal to the dataset size. This is because we do not count UNSAT generation results in the table.}
  \label{tab:res-z3-llm}

  \begin{tabular}{cccccc}
        \toprule
        No. of templates & No. generated by Z3 & No. generated by LLM & Z3 pass rate & LLM pass rate \\
        \midrule
        1 & 66 & 25 & 63.6\% & 48.0\%  \\
        2 & 74 & 23 & 73.0\% & 47.8\% \\
        3 & 76 & 24 & 69.7\% & 37.5\%\\
        4 & 71 & 19 & 70.4\% & 57.9\% \\
        5 & 66 & 29 & 68.2\% & 58.6\% \\
        \bottomrule
    \end{tabular}
\end{table}

Table~\ref{tab:res-z3-llm} compares the Z3 code generator with the LLM solver in the number and pass rate of generated path solutions. For all retrieval settings, the majority of solutions are generated by the Z3 code generator, and the pass rate of solutions generated from Z3 is significantly higher than those generated by the LLM solver. For example, when the number of retrieved templates is set to 2, the Z3 code generator generates 74 solutions, which is more than three times of the solutions generated by the LLM solver. Besides, the Z3 code generator reaches a high pass rate of 73\%. This shows the potential of building a symbolic execution engine fully powered by LLM to solve the constraints that are within the ability of Z3.

We also perform an ablation to evaluate the performance of the Z3 code generator: we compare \method with a baseline in which we use the LLM solver to solve \textbf{all} paths. This baseline can be seen as a variant of SymPrompt \cite{ryan2024code}, where the execution path is used in prompts for test case generation. The main difference is that our baseline uses execution paths extracted from CFGs instead of ASTs in SymPrompt. Table~\ref{tab:res-llmolver} displays the results of the LLM solver on our
dataset. We evaluate two proprietary models, GPT-4o-mini and GPT-4o, for the LLM solver. We record the number of successfully solved paths and the pass rate of the paths that are generated by Z3 code generator/LLM solver in \method. We find that although directly using the LLM for constraint solving can produce more executable test cases, its successfully solved paths are less than \method. Although GPT-4o has a higher overall pass rate than \method, its pass rate on the subset of Z3-solved paths is still lower than \method. Moreover, our Z3 code generator is based on the weaker and more efficient GPT-4o-mini. This indicates that using LLMs to call the SMT solver can free the LLM from complex constraint solving, and achieve better pass rates.

\begin{table}
  \caption{The results of only using the LLM solver on our dataset.}
  \label{tab:res-llmolver}

  \begin{tabular}{cccccc}
        \toprule
        Approach & SAT & Execution pass & Path correct & Pass rate (Z3) & Pass rate (LLM)\\
        \midrule
        gpt-4o-mini & 110 & 108 & 55 & 55.4\% & 39.1\%\\
        gpt-4o & 110 & 110 & \textbf{71} & 64.9\% & \textbf{52.2}\%\\
        \method & 97 & 94 & 65 & \textbf{73.0}\% & 47.8\% \\
        \bottomrule
    \end{tabular}
\end{table}

\begin{tcolorbox}[size=title,breakable]
\textbf{Answer to RQ3:} \textcolor{black}{The Z3 code generator can generate Z3Py code that better solves path constraints than only relying on the LLM. Meanwhile, for constraints that are difficult to solve with Z3, the LLM solver is a complementation with strong potential.}
\end{tcolorbox}

\subsection{RQ4: Time and cost efficiency}
The \method is an LLM agent, and its core, the Z3 code generator, requires multiple-step generation with self-refine. So, it is important to measure its efficiency. We compare \method with only using the LLM solver (without the Z3 code generator) in time and money costs, and the results are shown in Table~\ref{tab:cost}. In this table, we record the total time of \method in solving all 111 path constraints in our dataset and the total cost of using the OpenAI API.

\begin{table}
  \caption{The total cost (and cost per sample) for path constraints solving in our dataset with \method and different LLMs. The time metrics are shown in seconds, and the money costs are shown in US dollars.}
  \label{tab:cost}

  \begin{tabular}{ccc}
        \toprule
        Approach & Time & Money\\
        \midrule
        \method & 2241 (20.2) & 0.61 (0.005) \\
        gpt-4o-mini & 495 (4.5) & 0.04 (0.0004)\\
        gpt-4o & 1189 (10.7) & 1.96 (0.018)\\
        \bottomrule
    \end{tabular}
\end{table}

On average, \method solves a single path with 20.2s, and the cost is only 0.005\$. With a complicated line-by-line generation pipeline, its time overhead is less than twice that of the GPT-4o solver, and the money cost is only one-third of that of GPT-4o. While achieving similar results to GPT-4o, the price of our \method is significantly lower, and the time efficiency is still acceptable. 

\begin{tcolorbox}[size=title,breakable]
\textbf{Answer to RQ4:} \textcolor{black}{Although the Z3 code generation pipeline is complicated, the time overhead of \method is not much longer than the basic GPT-4o. The money cost of \method is significantly lower than GPT-4o, due to the low cost of the backbone LLM GPT-4o-mini.}
\end{tcolorbox}

\section{Threats to Validity}
\subsection{Model validity}
The proposed approach \method is based on LLMs GPT-4o-mini and GPT-4o, while other models are not considered. However, we should be aware that these two models are widely used state-of-the-art LLMs, so the experiment results are representative, and the choice of these models is reasonable.

Another threat is related to the capability of \method. Currently, our approach only supports a part of the Python `list' operation and does not explicitly support other data structures, which may be inadequate for real-world complex programs. However, because \method supports non-trivial operations by providing Z3Py examples in the knowledge base, we can add support for other operations by adding new examples into the knowledge base. 
Also, the ability of the key module in \method: the Z3 code generator, is limited by the abilities of the Z3 solver. To mitigate this threat, we introduced an LLM solver to solve constraints that Z3 cannot solve. Improvements on the LLM solver for stronger reasoning abilities are left as future work.

\subsection{Data validity}
In this paper, we have only evaluated \method on a subset of LeetCode problem solutions. As the first step of integrating LLMs with an SMT solver for symbolic execution, we should first evaluate simple, restricted programs and understand the behavior of \method on these data before we move on to more complicated, real-world data.

\section{Related Work}
\subsection{Symbolic execution}
Symbolic execution was first proposed to test whether a piece of software can violate certain properties \cite{baldoni2018survey}. Symbolic execution has been successful in many tasks, such as test case generation \cite{king1976symbolic}, inferring program invariants \cite{liu2023towards}, and vulnerability detection \cite{wang2024efficiently}. However, symbolic execution is known to suffer from many limitations, including path explosion and difficulties in solving non-linear constraints or external API calls. To address these limitations, one of the most adopted techniques is concolic execution \cite{sen2007concolic}. Instead of only executing with symbolic values, concolic execution executes the program with concrete values, collects path constraints, and explores other paths by negating constraints. Currently, many successful symbolic execution engines are based on concolic execution, including KLEE \cite{cadar2008klee}, SPF \cite{puasuareanu2010symbolic}, and Triton \cite{saudel2015triton}. Most symbolic execution engines use SMT solvers, such as Z3 \cite{de2008z3}, to solve path constraints.

Although existing symbolic execution engines have enabled symbolic analysis in many languages (e.g., C, Java, LLVM), few of them can support Python. We assume that the lack of Python symbolic execution engines is due to the unique features of Python: dynamic typing and data structures with flexible operations. Although several previous works \cite{bruni2011peer, bucur2014prototyping, ball2015deconstructing} tried to perform symbolic execution on Python, none of them have successfully transferred into established, well-maintained tools. For example, the Python symbolic execution tool PyExZ3 \cite{ball2015deconstructing} only supports simple data types (no support for `list') and has been deprecated. The most popular Python symbolic execution engine, CrossHair \footnote{https://github.com/pschanely/CrossHair} (with 900+ Github stars), is maintained by independent developers. We did not compare with CrossHair in our paper because it uses concolic execution and does not support pure-symbolic constraint solving. Another limitation of CrossHair is that it requires that the input arguments in the function under test must have explicit type annotations.

\subsection{LLM for test case generation}
Recently, software testing researchers have applied LLM for test case generation in different domains \cite{wang2024software}. One of the most discussed domains in LLM-aided software testing is unit test case generation \cite{schafer2023empirical, chen2024chatunitest, yuan2024evaluating, lemieux2023codamosa, ryan2024code}, which aims to generate test cases separately for individual software components. For example, ChatUniTest \cite{chen2024chatunitest} and ChatTester \cite{yuan2024evaluating} aim to leverage Openai GPT models to generate correct, runnable test cases for Java methods, while TrickyBugs \cite{liu2024trickybugs}, TestEval \cite{wang2024testeval} and SWT-Bench \cite{mundler2024code} built benchmark datasets to facilitate researches on generating bug-triggering or high-coverage test cases. There have also been some recent empirical studies \cite{yang2024empirical, ouedraogo2024large} which discussed the correctness, coverage, and bug-finding ability of LLM-generated test cases with different prompt designs. Symprompt \cite{ryan2024code} tried to combine LLM-based test case generation with symbolic analysis by using program execution paths to prompt LLMs. Although the overall idea of Symprompt resembles symbolic execution, it did not make additional efforts to improve the LLM's ability to solve path constraints. Instead, the execution path is provided to the LLM in one single prompt, which may hinder the LLM from solving the constraints with complex reasoning.

There are also other works that tried to combine LLMs with traditional program analysis tools or software testing techniques for test case generation. For example, CodaMosa \cite{lemieux2023codamosa} improves test coverage by integrating LLMs with Pynguin \cite{lukasczyk2022pynguin}, a search-based software testing tool. When Pynguin gets stuck in a coverage plateau, CodeMosa calls an LLM to generate new test case seeds so that Pynguin can generate new mutants on this seed and further improve test coverage. TELPA \cite{yang2024enhancing} leverages control flow graph analysis to extract method invocation sequences for guiding LLM-generated test cases to enter hard-to-cover branches. HITS \cite{wang2024hits} applies program slicing on Java methods under test and improves test coverage by generating test cases for different slices.


\section{Conclusion}
This paper proposes \method, a symbolic execution engine prototype which is powered by an LLM agent. It improves the solving ability of a basic symbolic execution engine by leveraging LLMs to call an SMT solver Z3 for constraint solving. Our proposed code generation approach allows \method to solve constraints on complex data types, e.g., Python lists, using the Z3 solver. If Z3 cannot solve the constraint, it turns to LLMs for solving. \method enables Z3 code generation for complex Python statements with a meticulously designed retrieval knowledge base and a self-refine mechanism. In this paper, we focus on supporting Python lists when building the knowledge base. Our experiments show that \method is capable of generating correct Z3 code for long path constraints in LeetCode programs. Moreover, we find that integrating a cost-efficient LLM with Z3 can beat the most state-of-the-art LLM in constraint solving for Python symbolic execution.


\bibliographystyle{ACM-Reference-Format}
\bibliography{sample-base}


\begin{thebibliography}{38}


\ifx \showCODEN    \undefined \def \showCODEN     #1{\unskip}     \fi
\ifx \showDOI      \undefined \def \showDOI       #1{#1}\fi
\ifx \showISBNx    \undefined \def \showISBNx     #1{\unskip}     \fi
\ifx \showISBNxiii \undefined \def \showISBNxiii  #1{\unskip}     \fi
\ifx \showISSN     \undefined \def \showISSN      #1{\unskip}     \fi
\ifx \showLCCN     \undefined \def \showLCCN      #1{\unskip}     \fi
\ifx \shownote     \undefined \def \shownote      #1{#1}          \fi
\ifx \showarticletitle \undefined \def \showarticletitle #1{#1}   \fi
\ifx \showURL      \undefined \def \showURL       {\relax}        \fi
\providecommand\bibfield[2]{#2}
\providecommand\bibinfo[2]{#2}
\providecommand\natexlab[1]{#1}
\providecommand\showeprint[2][]{arXiv:#2}

\bibitem[Baldoni et~al\mbox{.}(2018)]%
        {baldoni2018survey}
\bibfield{author}{\bibinfo{person}{Roberto Baldoni}, \bibinfo{person}{Emilio Coppa}, \bibinfo{person}{Daniele~Cono D’elia}, \bibinfo{person}{Camil Demetrescu}, {and} \bibinfo{person}{Irene Finocchi}.} \bibinfo{year}{2018}\natexlab{}.
\newblock \showarticletitle{A survey of symbolic execution techniques}.
\newblock \bibinfo{journal}{\emph{ACM Computing Surveys (CSUR)}} \bibinfo{volume}{51}, \bibinfo{number}{3} (\bibinfo{year}{2018}), \bibinfo{pages}{1--39}.
\newblock


\bibitem[Ball and Daniel(2015)]%
        {ball2015deconstructing}
\bibfield{author}{\bibinfo{person}{Thomas Ball} {and} \bibinfo{person}{Jakub Daniel}.} \bibinfo{year}{2015}\natexlab{}.
\newblock \showarticletitle{Deconstructing dynamic symbolic execution}.
\newblock In \bibinfo{booktitle}{\emph{Dependable Software Systems Engineering}}. \bibinfo{publisher}{IOS Press}, \bibinfo{pages}{26--41}.
\newblock
\urldef\tempurl%
\url{https://github.com/thomasjball/PyExZ3}
\showURL{%
\tempurl}


\bibitem[Bruni et~al\mbox{.}(2011)]%
        {bruni2011peer}
\bibfield{author}{\bibinfo{person}{Alessandro~Disney Bruni}, \bibinfo{person}{Tim Disney}, {and} \bibinfo{person}{Cormac Flanagan}.} \bibinfo{year}{2011}\natexlab{}.
\newblock \showarticletitle{A peer architecture for lightweight symbolic execution}.
\newblock \bibinfo{journal}{\emph{Universidad de California, Santa Cruz}} (\bibinfo{year}{2011}).
\newblock


\bibitem[Bucur et~al\mbox{.}(2014)]%
        {bucur2014prototyping}
\bibfield{author}{\bibinfo{person}{Stefan Bucur}, \bibinfo{person}{Johannes Kinder}, {and} \bibinfo{person}{George Candea}.} \bibinfo{year}{2014}\natexlab{}.
\newblock \showarticletitle{Prototyping symbolic execution engines for interpreted languages}. In \bibinfo{booktitle}{\emph{Proceedings of the 19th international conference on Architectural support for programming languages and operating systems}}. \bibinfo{pages}{239--254}.
\newblock


\bibitem[Cadar et~al\mbox{.}(2008)]%
        {cadar2008klee}
\bibfield{author}{\bibinfo{person}{Cristian Cadar}, \bibinfo{person}{Daniel Dunbar}, \bibinfo{person}{Dawson~R Engler}, {et~al\mbox{.}}} \bibinfo{year}{2008}\natexlab{}.
\newblock \showarticletitle{Klee: unassisted and automatic generation of high-coverage tests for complex systems programs.}. In \bibinfo{booktitle}{\emph{OSDI}}, Vol.~\bibinfo{volume}{8}. \bibinfo{pages}{209--224}.
\newblock


\bibitem[Chen et~al\mbox{.}(2024)]%
        {chen2024chatunitest}
\bibfield{author}{\bibinfo{person}{Yinghao Chen}, \bibinfo{person}{Zehao Hu}, \bibinfo{person}{Chen Zhi}, \bibinfo{person}{Junxiao Han}, \bibinfo{person}{Shuiguang Deng}, {and} \bibinfo{person}{Jianwei Yin}.} \bibinfo{year}{2024}\natexlab{}.
\newblock \showarticletitle{Chatunitest: A framework for llm-based test generation}. In \bibinfo{booktitle}{\emph{Companion Proceedings of the 32nd ACM International Conference on the Foundations of Software Engineering}}. \bibinfo{pages}{572--576}.
\newblock


\bibitem[Cobbe et~al\mbox{.}(2021)]%
        {cobbe2021training}
\bibfield{author}{\bibinfo{person}{Karl Cobbe}, \bibinfo{person}{Vineet Kosaraju}, \bibinfo{person}{Mohammad Bavarian}, \bibinfo{person}{Mark Chen}, \bibinfo{person}{Heewoo Jun}, \bibinfo{person}{Lukasz Kaiser}, \bibinfo{person}{Matthias Plappert}, \bibinfo{person}{Jerry Tworek}, \bibinfo{person}{Jacob Hilton}, \bibinfo{person}{Reiichiro Nakano}, {et~al\mbox{.}}} \bibinfo{year}{2021}\natexlab{}.
\newblock \showarticletitle{Training verifiers to solve math word problems}.
\newblock \bibinfo{journal}{\emph{arXiv preprint arXiv:2110.14168}} (\bibinfo{year}{2021}).
\newblock


\bibitem[De~Moura and Bj{\o}rner(2008)]%
        {de2008z3}
\bibfield{author}{\bibinfo{person}{Leonardo De~Moura} {and} \bibinfo{person}{Nikolaj Bj{\o}rner}.} \bibinfo{year}{2008}\natexlab{}.
\newblock \showarticletitle{Z3: An efficient SMT solver}. In \bibinfo{booktitle}{\emph{International conference on Tools and Algorithms for the Construction and Analysis of Systems}}. Springer, \bibinfo{pages}{337--340}.
\newblock


\bibitem[Galeotti et~al\mbox{.}(2013)]%
        {galeotti2013improving}
\bibfield{author}{\bibinfo{person}{Juan~Pablo Galeotti}, \bibinfo{person}{Gordon Fraser}, {and} \bibinfo{person}{Andrea Arcuri}.} \bibinfo{year}{2013}\natexlab{}.
\newblock \showarticletitle{Improving search-based test suite generation with dynamic symbolic execution}. In \bibinfo{booktitle}{\emph{2013 ieee 24th international symposium on software reliability engineering (issre)}}. IEEE, \bibinfo{pages}{360--369}.
\newblock


\bibitem[King(1976)]%
        {king1976symbolic}
\bibfield{author}{\bibinfo{person}{James~C King}.} \bibinfo{year}{1976}\natexlab{}.
\newblock \showarticletitle{Symbolic execution and program testing}.
\newblock \bibinfo{journal}{\emph{Commun. ACM}} \bibinfo{volume}{19}, \bibinfo{number}{7} (\bibinfo{year}{1976}), \bibinfo{pages}{385--394}.
\newblock


\bibitem[Lemieux et~al\mbox{.}(2023)]%
        {lemieux2023codamosa}
\bibfield{author}{\bibinfo{person}{Caroline Lemieux}, \bibinfo{person}{Jeevana~Priya Inala}, \bibinfo{person}{Shuvendu~K Lahiri}, {and} \bibinfo{person}{Siddhartha Sen}.} \bibinfo{year}{2023}\natexlab{}.
\newblock \showarticletitle{Codamosa: Escaping coverage plateaus in test generation with pre-trained large language models}. In \bibinfo{booktitle}{\emph{2023 IEEE/ACM 45th International Conference on Software Engineering (ICSE)}}. IEEE, \bibinfo{pages}{919--931}.
\newblock


\bibitem[Li et~al\mbox{.}(2024)]%
        {li2024holistic}
\bibfield{author}{\bibinfo{person}{Penghui Li}, \bibinfo{person}{Wei Meng}, \bibinfo{person}{Mingxue Zhang}, \bibinfo{person}{Chenlin Wang}, {and} \bibinfo{person}{Changhua Luo}.} \bibinfo{year}{2024}\natexlab{}.
\newblock \showarticletitle{Holistic Concolic Execution for Dynamic Web Applications via Symbolic Interpreter Analysis}. In \bibinfo{booktitle}{\emph{Proceedings of the 45th IEEE Symposium on Security and Privacy (Oakland). San Francisco, CA, USA}}.
\newblock


\bibitem[Li et~al\mbox{.}(2016)]%
        {li2016symbolic}
\bibfield{author}{\bibinfo{person}{Xin Li}, \bibinfo{person}{Yongjuan Liang}, \bibinfo{person}{Hong Qian}, \bibinfo{person}{Yi-Qi Hu}, \bibinfo{person}{Lei Bu}, \bibinfo{person}{Yang Yu}, \bibinfo{person}{Xin Chen}, {and} \bibinfo{person}{Xuandong Li}.} \bibinfo{year}{2016}\natexlab{}.
\newblock \showarticletitle{Symbolic execution of complex program driven by machine learning based constraint solving}. In \bibinfo{booktitle}{\emph{Proceedings of the 31st IEEE/ACM International Conference on Automated Software Engineering}}. \bibinfo{pages}{554--559}.
\newblock


\bibitem[Liu et~al\mbox{.}(2023)]%
        {liu2023towards}
\bibfield{author}{\bibinfo{person}{Chang Liu}, \bibinfo{person}{Xiwei Wu}, \bibinfo{person}{Yuan Feng}, \bibinfo{person}{Qinxiang Cao}, {and} \bibinfo{person}{Junchi Yan}.} \bibinfo{year}{2023}\natexlab{}.
\newblock \showarticletitle{Towards General Loop Invariant Generation via Coordinating Symbolic Execution and Large Language Models}.
\newblock \bibinfo{journal}{\emph{arXiv preprint arXiv:2311.10483}} (\bibinfo{year}{2023}).
\newblock


\bibitem[Liu et~al\mbox{.}(2024)]%
        {liu2024trickybugs}
\bibfield{author}{\bibinfo{person}{Kaibo Liu}, \bibinfo{person}{Yudong Han}, \bibinfo{person}{Yiyang Liu}, \bibinfo{person}{Zhenpeng Chen}, \bibinfo{person}{Jie~M Zhang}, \bibinfo{person}{Federica Sarro}, \bibinfo{person}{Gang Huang}, {and} \bibinfo{person}{Yun Ma}.} \bibinfo{year}{2024}\natexlab{}.
\newblock \showarticletitle{TrickyBugs: A Dataset of Corner-case Bugs in Plausible Programs}. In \bibinfo{booktitle}{\emph{Proceedings of the 21st International Conference on Mining Software Repositories}}. \bibinfo{pages}{113--117}.
\newblock


\bibitem[Lukasczyk and Fraser(2022)]%
        {lukasczyk2022pynguin}
\bibfield{author}{\bibinfo{person}{Stephan Lukasczyk} {and} \bibinfo{person}{Gordon Fraser}.} \bibinfo{year}{2022}\natexlab{}.
\newblock \showarticletitle{Pynguin: Automated unit test generation for python}. In \bibinfo{booktitle}{\emph{Proceedings of the ACM/IEEE 44th International Conference on Software Engineering: Companion Proceedings}}. \bibinfo{pages}{168--172}.
\newblock


\bibitem[M{\"u}ndler et~al\mbox{.}(2024)]%
        {mundler2024code}
\bibfield{author}{\bibinfo{person}{Niels M{\"u}ndler}, \bibinfo{person}{Mark~Niklas M{\"u}ller}, \bibinfo{person}{Jingxuan He}, {and} \bibinfo{person}{Martin Vechev}.} \bibinfo{year}{2024}\natexlab{}.
\newblock \showarticletitle{Code Agents are State of the Art Software Testers}.
\newblock \bibinfo{journal}{\emph{arXiv preprint arXiv:2406.12952}} (\bibinfo{year}{2024}).
\newblock


\bibitem[Ou{\'e}draogo et~al\mbox{.}(2024)]%
        {ouedraogo2024large}
\bibfield{author}{\bibinfo{person}{Wendk{\^u}uni~C Ou{\'e}draogo}, \bibinfo{person}{Kader Kabor{\'e}}, \bibinfo{person}{Haoye Tian}, \bibinfo{person}{Yewei Song}, \bibinfo{person}{Anil Koyuncu}, \bibinfo{person}{Jacques Klein}, \bibinfo{person}{David Lo}, {and} \bibinfo{person}{Tegawend{\'e}~F Bissyand{\'e}}.} \bibinfo{year}{2024}\natexlab{}.
\newblock \showarticletitle{Large-scale, Independent and Comprehensive study of the power of LLMs for test case generation}.
\newblock \bibinfo{journal}{\emph{arXiv preprint arXiv:2407.00225}} (\bibinfo{year}{2024}).
\newblock


\bibitem[Pan et~al\mbox{.}(2023)]%
        {pan2023logic}
\bibfield{author}{\bibinfo{person}{Liangming Pan}, \bibinfo{person}{Alon Albalak}, \bibinfo{person}{Xinyi Wang}, {and} \bibinfo{person}{William Wang}.} \bibinfo{year}{2023}\natexlab{}.
\newblock \showarticletitle{Logic-LM: Empowering Large Language Models with Symbolic Solvers for Faithful Logical Reasoning}. In \bibinfo{booktitle}{\emph{Findings of the Association for Computational Linguistics: EMNLP 2023}}. \bibinfo{pages}{3806--3824}.
\newblock


\bibitem[P{\u{a}}s{\u{a}}reanu and Rungta(2010)]%
        {puasuareanu2010symbolic}
\bibfield{author}{\bibinfo{person}{Corina~S P{\u{a}}s{\u{a}}reanu} {and} \bibinfo{person}{Neha Rungta}.} \bibinfo{year}{2010}\natexlab{}.
\newblock \showarticletitle{Symbolic PathFinder: symbolic execution of Java bytecode}. In \bibinfo{booktitle}{\emph{Proceedings of the 25th IEEE/ACM International Conference on Automated Software Engineering}}. \bibinfo{pages}{179--180}.
\newblock


\bibitem[Reimers and Gurevych(2019)]%
        {reimers-2019-sentence-bert}
\bibfield{author}{\bibinfo{person}{Nils Reimers} {and} \bibinfo{person}{Iryna Gurevych}.} \bibinfo{year}{2019}\natexlab{}.
\newblock \showarticletitle{Sentence-BERT: Sentence Embeddings using Siamese BERT-Networks}. In \bibinfo{booktitle}{\emph{Proceedings of the 2019 Conference on Empirical Methods in Natural Language Processing}}. \bibinfo{publisher}{Association for Computational Linguistics}.
\newblock
\urldef\tempurl%
\url{https://arxiv.org/abs/1908.10084}
\showURL{%
\tempurl}


\bibitem[Ryan et~al\mbox{.}(2024)]%
        {ryan2024code}
\bibfield{author}{\bibinfo{person}{Gabriel Ryan}, \bibinfo{person}{Siddhartha Jain}, \bibinfo{person}{Mingyue Shang}, \bibinfo{person}{Shiqi Wang}, \bibinfo{person}{Xiaofei Ma}, \bibinfo{person}{Murali~Krishna Ramanathan}, {and} \bibinfo{person}{Baishakhi Ray}.} \bibinfo{year}{2024}\natexlab{}.
\newblock \showarticletitle{Code-Aware Prompting: A Study of Coverage-Guided Test Generation in Regression Setting using LLM}.
\newblock \bibinfo{journal}{\emph{Proceedings of the ACM on Software Engineering}} \bibinfo{volume}{1}, \bibinfo{number}{FSE} (\bibinfo{year}{2024}), \bibinfo{pages}{951--971}.
\newblock


\bibitem[Saudel and Salwan(2015)]%
        {saudel2015triton}
\bibfield{author}{\bibinfo{person}{Florent Saudel} {and} \bibinfo{person}{Jonathan Salwan}.} \bibinfo{year}{2015}\natexlab{}.
\newblock \showarticletitle{Triton: A dynamic symbolic execution framework}. In \bibinfo{booktitle}{\emph{Symposium sur la s{\'e}curit{\'e} des technologies de l’information et des communications, SSTIC, France, Rennes}}. \bibinfo{pages}{31--54}.
\newblock


\bibitem[Sch{\"a}fer et~al\mbox{.}(2023)]%
        {schafer2023empirical}
\bibfield{author}{\bibinfo{person}{Max Sch{\"a}fer}, \bibinfo{person}{Sarah Nadi}, \bibinfo{person}{Aryaz Eghbali}, {and} \bibinfo{person}{Frank Tip}.} \bibinfo{year}{2023}\natexlab{}.
\newblock \showarticletitle{An empirical evaluation of using large language models for automated unit test generation}.
\newblock \bibinfo{journal}{\emph{IEEE Transactions on Software Engineering}} (\bibinfo{year}{2023}).
\newblock


\bibitem[Sen(2007)]%
        {sen2007concolic}
\bibfield{author}{\bibinfo{person}{Koushik Sen}.} \bibinfo{year}{2007}\natexlab{}.
\newblock \showarticletitle{Concolic testing}. In \bibinfo{booktitle}{\emph{Proceedings of the 22nd IEEE/ACM international conference on Automated software engineering}}. \bibinfo{pages}{571--572}.
\newblock


\bibitem[Shoshitaishvili et~al\mbox{.}(2016)]%
        {shoshitaishvili2016sok}
\bibfield{author}{\bibinfo{person}{Yan Shoshitaishvili}, \bibinfo{person}{Ruoyu Wang}, \bibinfo{person}{Christopher Salls}, \bibinfo{person}{Nick Stephens}, \bibinfo{person}{Mario Polino}, \bibinfo{person}{Andrew Dutcher}, \bibinfo{person}{John Grosen}, \bibinfo{person}{Siji Feng}, \bibinfo{person}{Christophe Hauser}, \bibinfo{person}{Christopher Kruegel}, {et~al\mbox{.}}} \bibinfo{year}{2016}\natexlab{}.
\newblock \showarticletitle{Sok:(state of) the art of war: Offensive techniques in binary analysis}. In \bibinfo{booktitle}{\emph{2016 IEEE symposium on security and privacy (SP)}}. IEEE, \bibinfo{pages}{138--157}.
\newblock


\bibitem[Srivastava et~al\mbox{.}(2022)]%
        {srivastava2022beyond}
\bibfield{author}{\bibinfo{person}{Aarohi Srivastava}, \bibinfo{person}{Abhinav Rastogi}, \bibinfo{person}{Abhishek Rao}, \bibinfo{person}{Abu Awal~Md Shoeb}, \bibinfo{person}{Abubakar Abid}, \bibinfo{person}{Adam Fisch}, \bibinfo{person}{Adam~R Brown}, \bibinfo{person}{Adam Santoro}, \bibinfo{person}{Aditya Gupta}, \bibinfo{person}{Adri{\`a} Garriga-Alonso}, {et~al\mbox{.}}} \bibinfo{year}{2022}\natexlab{}.
\newblock \showarticletitle{Beyond the imitation game: Quantifying and extrapolating the capabilities of language models}.
\newblock \bibinfo{journal}{\emph{arXiv preprint arXiv:2206.04615}} (\bibinfo{year}{2022}).
\newblock


\bibitem[Wang et~al\mbox{.}(2024e)]%
        {wang2024dataflow}
\bibfield{author}{\bibinfo{person}{Chengpeng Wang}, \bibinfo{person}{Wuqi Zhang}, \bibinfo{person}{Zian Su}, \bibinfo{person}{Xiangzhe Xu}, \bibinfo{person}{Xiaoheng Xie}, {and} \bibinfo{person}{Xiangyu Zhang}.} \bibinfo{year}{2024}\natexlab{e}.
\newblock \showarticletitle{When Dataflow Analysis Meets Large Language Models}.
\newblock \bibinfo{journal}{\emph{arXiv preprint arXiv:2402.10754}} (\bibinfo{year}{2024}).
\newblock


\bibitem[Wang et~al\mbox{.}(2024b)]%
        {wang2024software}
\bibfield{author}{\bibinfo{person}{Junjie Wang}, \bibinfo{person}{Yuchao Huang}, \bibinfo{person}{Chunyang Chen}, \bibinfo{person}{Zhe Liu}, \bibinfo{person}{Song Wang}, {and} \bibinfo{person}{Qing Wang}.} \bibinfo{year}{2024}\natexlab{b}.
\newblock \showarticletitle{Software testing with large language models: Survey, landscape, and vision}.
\newblock \bibinfo{journal}{\emph{IEEE Transactions on Software Engineering}} (\bibinfo{year}{2024}).
\newblock


\bibitem[Wang et~al\mbox{.}(2024d)]%
        {wang2024testeval}
\bibfield{author}{\bibinfo{person}{Wenhan Wang}, \bibinfo{person}{Chenyuan Yang}, \bibinfo{person}{Zhijie Wang}, \bibinfo{person}{Yuheng Huang}, \bibinfo{person}{Zhaoyang Chu}, \bibinfo{person}{Da Song}, \bibinfo{person}{Lingming Zhang}, \bibinfo{person}{An~Ran Chen}, {and} \bibinfo{person}{Lei Ma}.} \bibinfo{year}{2024}\natexlab{d}.
\newblock \showarticletitle{TESTEVAL: Benchmarking Large Language Models for Test Case Generation}.
\newblock \bibinfo{journal}{\emph{arXiv preprint arXiv:2406.04531}} (\bibinfo{year}{2024}).
\newblock


\bibitem[Wang et~al\mbox{.}(2024a)]%
        {wang2024efficiently}
\bibfield{author}{\bibinfo{person}{Zexu Wang}, \bibinfo{person}{Jiachi Chen}, \bibinfo{person}{Yanlin Wang}, \bibinfo{person}{Yu Zhang}, \bibinfo{person}{Weizhe Zhang}, {and} \bibinfo{person}{Zibin Zheng}.} \bibinfo{year}{2024}\natexlab{a}.
\newblock \showarticletitle{Efficiently detecting reentrancy vulnerabilities in complex smart contracts}.
\newblock \bibinfo{journal}{\emph{Proceedings of the ACM on Software Engineering}} \bibinfo{volume}{1}, \bibinfo{number}{FSE} (\bibinfo{year}{2024}), \bibinfo{pages}{161--181}.
\newblock


\bibitem[Wang et~al\mbox{.}(2024c)]%
        {wang2024hits}
\bibfield{author}{\bibinfo{person}{Zejun Wang}, \bibinfo{person}{Kaibo Liu}, \bibinfo{person}{Ge Li}, {and} \bibinfo{person}{Zhi Jin}.} \bibinfo{year}{2024}\natexlab{c}.
\newblock \showarticletitle{HITS: High-coverage LLM-based Unit Test Generation via Method Slicing}.
\newblock \bibinfo{journal}{\emph{arXiv preprint arXiv:2408.11324}} (\bibinfo{year}{2024}).
\newblock


\bibitem[Wu et~al\mbox{.}(2024)]%
        {wu2024autogen}
\bibfield{author}{\bibinfo{person}{Qingyun Wu}, \bibinfo{person}{Gagan Bansal}, \bibinfo{person}{Jieyu Zhang}, \bibinfo{person}{Yiran Wu}, \bibinfo{person}{Beibin Li}, \bibinfo{person}{Erkang Zhu}, \bibinfo{person}{Li Jiang}, \bibinfo{person}{Xiaoyun Zhang}, \bibinfo{person}{Shaokun Zhang}, \bibinfo{person}{Jiale Liu}, {et~al\mbox{.}}} \bibinfo{year}{2024}\natexlab{}.
\newblock \showarticletitle{AutoGen: Enabling Next-Gen LLM Applications via Multi-Agent Conversation}. In \bibinfo{booktitle}{\emph{ICLR 2024 Workshop on Large Language Model (LLM) Agents}}.
\newblock


\bibitem[Yang et~al\mbox{.}(2024a)]%
        {yang2024enhancing}
\bibfield{author}{\bibinfo{person}{Chen Yang}, \bibinfo{person}{Junjie Chen}, \bibinfo{person}{Bin Lin}, \bibinfo{person}{Jianyi Zhou}, {and} \bibinfo{person}{Ziqi Wang}.} \bibinfo{year}{2024}\natexlab{a}.
\newblock \showarticletitle{Enhancing LLM-based Test Generation for Hard-to-Cover Branches via Program Analysis}.
\newblock \bibinfo{journal}{\emph{arXiv preprint arXiv:2404.04966}} (\bibinfo{year}{2024}).
\newblock


\bibitem[Yang et~al\mbox{.}(2024b)]%
        {yang2024empirical}
\bibfield{author}{\bibinfo{person}{Lin Yang}, \bibinfo{person}{Chen Yang}, \bibinfo{person}{Shutao Gao}, \bibinfo{person}{Weijing Wang}, \bibinfo{person}{Bo Wang}, \bibinfo{person}{Qihao Zhu}, \bibinfo{person}{Xiao Chu}, \bibinfo{person}{Jianyi Zhou}, \bibinfo{person}{Guangtai Liang}, \bibinfo{person}{Qianxiang Wang}, {et~al\mbox{.}}} \bibinfo{year}{2024}\natexlab{b}.
\newblock \showarticletitle{An Empirical Study of Unit Test Generation with Large Language Models}.
\newblock \bibinfo{journal}{\emph{arXiv preprint arXiv:2406.18181}} (\bibinfo{year}{2024}).
\newblock


\bibitem[Ye et~al\mbox{.}(2023)]%
        {ye2024satlm}
\bibfield{author}{\bibinfo{person}{Xi Ye}, \bibinfo{person}{Qiaochu Chen}, \bibinfo{person}{Isil Dillig}, {and} \bibinfo{person}{Greg Durrett}.} \bibinfo{year}{2023}\natexlab{}.
\newblock \showarticletitle{Satlm: Satisfiability-aided language models using declarative prompting}.
\newblock \bibinfo{journal}{\emph{Advances in Neural Information Processing Systems}}  \bibinfo{volume}{36} (\bibinfo{year}{2023}).
\newblock


\bibitem[Yuan et~al\mbox{.}(2024)]%
        {yuan2024evaluating}
\bibfield{author}{\bibinfo{person}{Zhiqiang Yuan}, \bibinfo{person}{Mingwei Liu}, \bibinfo{person}{Shiji Ding}, \bibinfo{person}{Kaixin Wang}, \bibinfo{person}{Yixuan Chen}, \bibinfo{person}{Xin Peng}, {and} \bibinfo{person}{Yiling Lou}.} \bibinfo{year}{2024}\natexlab{}.
\newblock \showarticletitle{Evaluating and improving chatgpt for unit test generation}.
\newblock \bibinfo{journal}{\emph{Proceedings of the ACM on Software Engineering}} \bibinfo{volume}{1}, \bibinfo{number}{FSE} (\bibinfo{year}{2024}), \bibinfo{pages}{1703--1726}.
\newblock


\bibitem[Zeller et~al\mbox{.}(2019)]%
        {zeller2019fuzzing}
\bibfield{author}{\bibinfo{person}{Andreas Zeller}, \bibinfo{person}{Rahul Gopinath}, \bibinfo{person}{Marcel B{\"o}hme}, \bibinfo{person}{Gordon Fraser}, {and} \bibinfo{person}{Christian Holler}.} \bibinfo{year}{2019}\natexlab{}.
\newblock \bibinfo{title}{The fuzzing book}.
\newblock
\newblock


\end{thebibliography}

\end{document}